\documentclass{article}
\usepackage{spconf,amsmath,graphicx,hyperref}
\usepackage{makecell}
\usepackage{amssymb}
\usepackage{subcaption}
\ninept

\title{Noise-to-Notes: Diffusion-based Generation and Refinement for Automatic Drum Transcription}

\name{Michael Yeung \qquad Keisuke Toyama \qquad Toya Teramoto \qquad Shusuke Takahashi \qquad Tamaki Kojima}
  
\address{Sony Group Corporation, Tokyo, Japan}
\begin{document}
%
\maketitle

\begin{abstract}
Automatic drum transcription (ADT) is traditionally formulated as a discriminative task to predict drum events from audio spectrograms. In this work, we redefine ADT as a conditional generative task and introduce Noise-to-Notes (N2N), a framework leveraging diffusion modeling to transform audio-conditioned Gaussian noise into drum events with associated velocities. This generative diffusion approach offers distinct advantages, including a flexible speed-accuracy trade-off and strong inpainting capabilities. However, the generation of binary onset and continuous velocity values presents a challenge for diffusion models, and to overcome this, we introduce an Annealed Pseudo-Huber loss to facilitate effective joint optimization. Finally, to augment low-level spectrogram features, we propose incorporating features extracted from music foundation models (MFMs), which capture high-level semantic information and enhance robustness to out-of-domain drum audio. Experimental results demonstrate that including MFM features significantly improves robustness and N2N establishes a new state-of-the-art performance across multiple ADT benchmarks.

\end{abstract}

\section{Introduction}
\label{sec:introduction}

Automatic drum transcription (ADT) is a subset of automatic music transcription, where the task is to derive symbolic musical representations from audio recordings. ADT primarily focuses on transcribing drum onset times, although the inclusion of velocity (dynamics) has also been shown to significantly enhance perceptual quality \cite{callender2020improving}. Despite the development of automatic methods to handle drum audio mixed with percussive or melodic instruments, consistent performance has only been achieved for simplified drum configurations \cite{wu2018review}. Discriminating between different drum components is challenging because spectrograms derived from drum audio lack clear harmonic structure, with significant overlap in time and frequency components among instruments \cite{wu2018review}. Moreover, the spectral characteristics of the same drum component may vary significantly depending on the sound sources and production methods used. Rather than directly using spectrogram features, music foundation models (MFMs) have been shown to provide effective features for various downstream music-related tasks \cite{li2023mert} and this could be leveraged to disentangle the signals from different drum components.

While generative models have not been explored in ADT, recent research in other classification tasks suggests that they can outperform discriminative approaches \cite{jaini2023intriguing}. The additional modeling of the data distribution could help mitigate overfitting, a significant issue in ADT. Here, we introduce the first generative model for ADT, Noise-to-Notes (N2N). We propose a novel diffusion objective to facilitate joint onset-velocity prediction, demonstrate strong inpainting capability and improve robustness by integrating spectrogram and MFM features, outperforming discriminative models and achieving state-of-the-art results across ADT benchmarks.




\section{Related works}
\label{sec:related_works}

\begin{figure}
  \centering
   \includegraphics[width=1.0\linewidth]{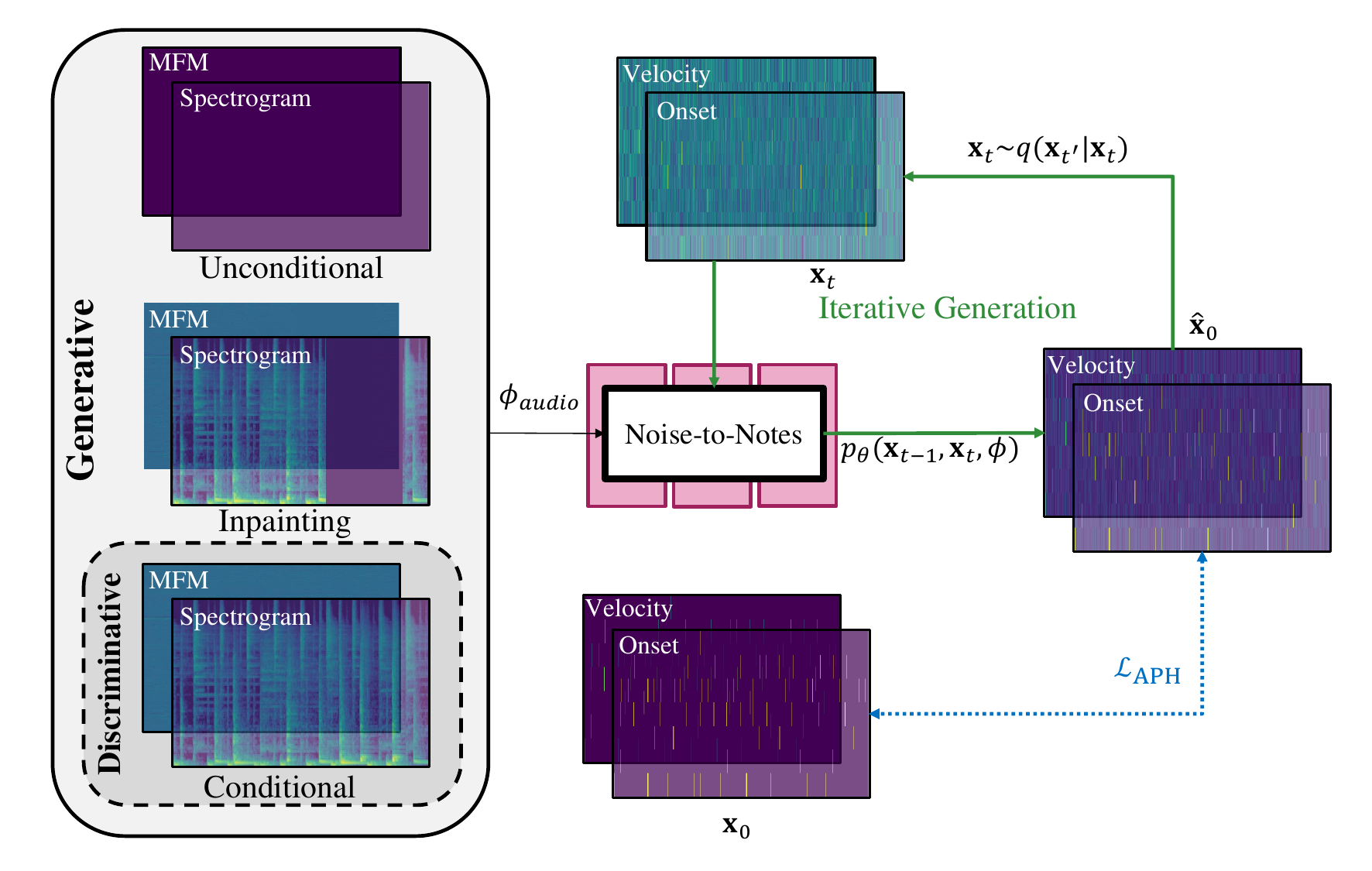}
   \caption{\textbf{Overview.} Noise-to-Notes is a diffusion model that transcribes drum audio. By reframing ADT as a generative task, N2N is capable of transcription with complete audio (conditional), partial audio (inpainting), and absence of audio (unconditional).}
   \label{fig:overview}
\end{figure}

Supervised deep learning-based approaches are the state-of-the-art for ADT \cite{wu2018review}. Specifically, convolutional recurrent neural networks (CRNNs) have remained the dominant architecture for ADT methods \cite{vogl2017drum,vogl2018towards,weber2025star}, which combines convolutional and recurrent layers for local acoustic and temporal modeling, respectively. Limited success has been achieved through architectural modifications \cite{zehren2023high}, and recent performance improvements can be mainly attributed to the curation of larger and more diverse datasets \cite{zehren2021adtof,wei2021improving}. 

Leveraging rhythm game annotations, Zehren et al. \cite{zehren2021adtof} created the ADTOF dataset, containing 114 hours of real annotated drum onsets and trained a CRNN to achieve state-of-the-art results across multiple ADT benchmarks. In addition to drum onsets, the E-GMD dataset \cite{callender2020improving} contains velocity annotations, and the associated Onset and Frames (OaF) Drums model, also based on CRNNs, achieved significantly better perceptual quality in human studies due to the inclusion of estimated velocities. While CRNNs remain the most successful approach in ADT, transformers offer more flexible time and frequency modeling and have become the preferred approach for the closely related automatic piano transcription task \cite{toyama2023automatic,yan2024scoring,10896775}. 


Beyond discriminative models, there has also been recent interest in modeling piano transcription as a generative task using diffusion models \cite{cheuk2023diffroll,10889579}. Diffusion models \cite{song2019generative,ho2020denoising} are a class of deep generative models that iteratively add noise to data and learn to reverse this process, enabling the generation of high-quality samples by progressively denoising from a stationary distribution. DiffRoll \cite{cheuk2023diffroll} was the first method to reframe music transcription as a generative task, and used a U-Net-based diffusion model to predict piano onsets from mel-spectrogram features. More recently, DR3M \cite{10889579} used a discrete denoising process with neigborhood attention to avoid modeling onsets as continuous variables, and trained an acoustic encoder to extract multi-scale mel-spectrogram features.

While modeling music transcription as a generative task has opened new possibilities for generation and refinement, neither DiffRoll nor DR3M reach the performance of discriminative models. Moreover, these methods focus on modeling onset, and the generative quality remains limited.

\section{Method}
\label{sec:method}

N2N is an audio-conditioned diffusion model for simultaneous drum onset and velocity transcription. An overview is shown in Fig. \ref{fig:overview}. 

Drum transcription is formulated as a frame-wise classification task, with the target transcription represented as $\textbf{x} \in \mathbb{R}^{F \times D \times 2}$, where $F$ is the number of frames and $D$ is the number of drum components. For each instrument at each frame, the ground truth annotations ($\text{\textbf{x}}_0$) consist of an onset $\in \{0,1\}$, indicating the presence of a drum hit, and associated velocities $\in [0,127]$ representing intensity. For training, we rescale both onset and velocity to $[-1,1]$.

Here, we reframe ADT in a diffusion framework, where a denoising network $\mathcal{D}_{\theta}(\text{\textbf{x}}_{\sigma_{t}};\sigma_t, \phi)$ is trained to recover a clean transcription from noisy transcriptions, conditioned on audio and timestep features ($\phi$). Concretely, using the original DDPM formulation \cite{ho2020denoising}, the forward process $p(\text{\textbf{x}}_{\sigma_{t}}; \sigma_t)$ is used to generate noisy targets by adding i.i.d. Gaussian noise ($\sigma$) to a clean transcription from a predefined, monotonically increasing sequence of noise levels $\sigma_0 = \sigma_{\text{max}} > \sigma_1 > ... > \sigma_N = \sigma_{\text{min}}$, such that the endpoint of the sequence corresponds to sampling from Gaussian noise at $\sigma_{\text{max}}$. During training, a randomly sampled noise level $t$ is used to generate noisy transcriptions $\text{\textbf{x}}_t \sim p(\text{\textbf{x}}_t;\sigma_t)$. The denoising network can then be trained by minimizing the mean-squared error (MSE) loss:

\begin{equation}
\mathbb{E}_{\text{\textbf{x}} \sim p_{\text{data}}(\text{\textbf{x}})} \mathbb{E}_{\epsilon, \sigma_t \sim p(\epsilon, \sigma_t)} \left[ \| \mathcal{D}_\theta(\text{\textbf{x}}_{\sigma_t}; \sigma_t, \phi) - \text{\textbf{x}} \|_2^2 \right].
\end{equation}

However, modeling onsets and velocities as continuous values poses an optimization challenge and we observe that using this standard denoising objective is suboptimal (Sec \ref{sec:ablation}). To avoid onset errors dominating the loss, we leverage the Pseudo-Huber loss \cite{charbonnier1997deterministic}, which has been used for training consistency models \cite{song2024improved} and offers a smooth interpolation between MSE and mean absolute error (MAE) objectives using a fixed constant $c$. However, we observe that fixing $c$ hinders optimization (Sec \ref{sec:ablation}), and instead we formulate an Annealed Pseudo-Huber loss ($\mathcal{L}_{\text{APH}}$):

\begin{equation}
\mathcal{L}_{\text{APH}}(\textbf{x},{\hat{\textbf{x}}}) =  \sqrt{\|\textbf{x} - \hat{\textbf{x}}\|_2^2 + c(t)^2} - c(t),
\end{equation}

where $c(t)$ is scheduled during training to shift the objective from MSE loss at the beginning, to MAE loss near the end of training. Empirically, we use $c(t) = (1 - \alpha_t) c_{\text{max}} + \alpha_t c_{\text{min}}$, where $c_{\text{max}} = 1$, $c_{\text{min}}=10^{-4}$, and $\alpha_t$ is linearly annealed from 0 to 1.

\begin{figure}[ht]
  \centering
   \includegraphics[width=1.0\linewidth]{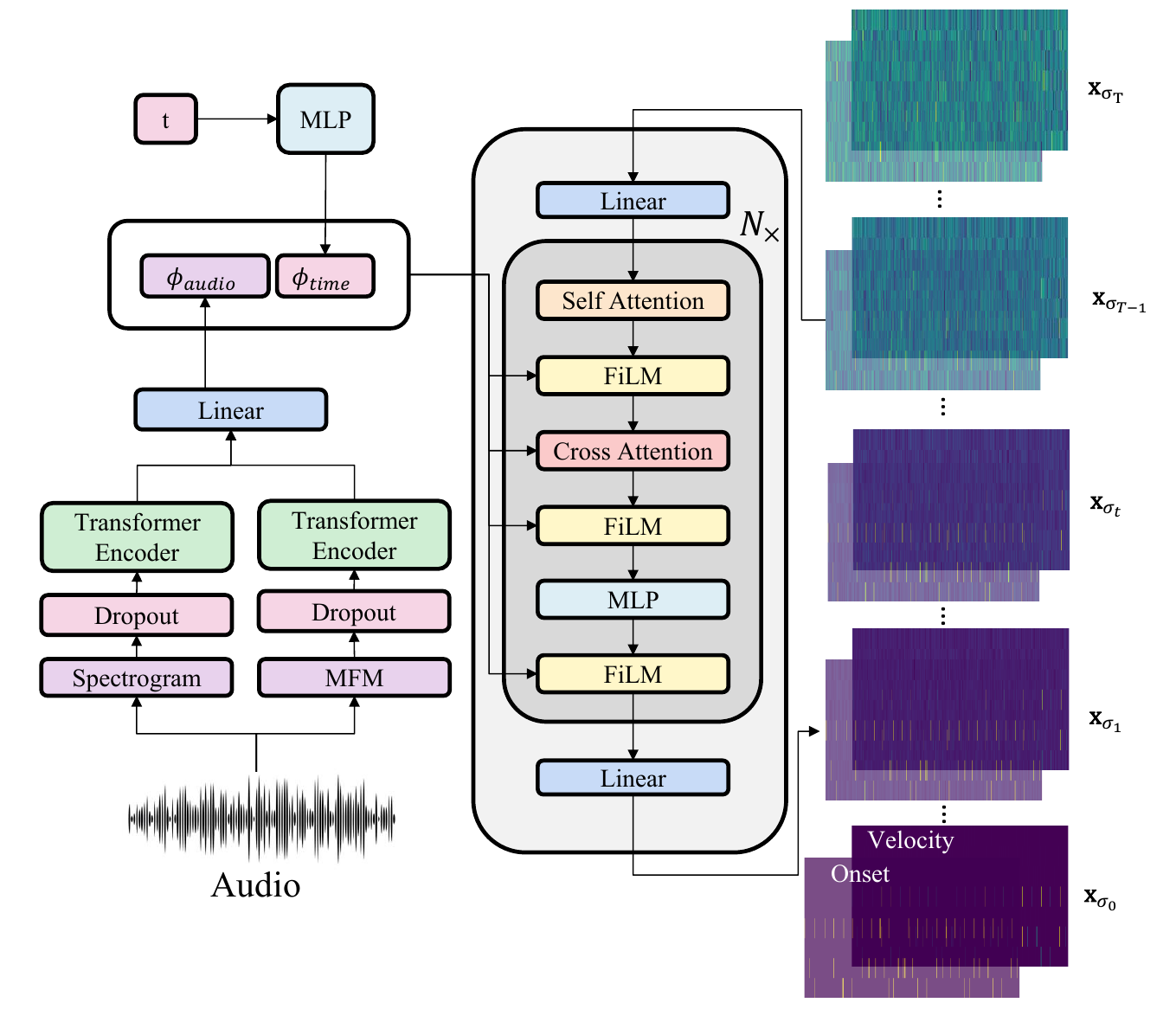}
   \caption{\textbf{Noise-to-Notes architecture}. N2N is an audio-conditioned transformer-based diffusion model. Drum audio features are extracted with a log mel-spectrogram and music foundation model. These features, combined with timestep ($\sigma_t$) information, modulate the decoder through cross attention and FiLM layers.}
   \label{fig:architecture}
\end{figure}

In practice, we used a more recent diffusion framework \cite{karras2022elucidating} that includes several modifications to improve generation efficiency and quality. Specifically, a lognormal noise schedule is used for training: $\ln \sigma \sim \mathcal{N}(P_{\text{mean}}, P_{\text{std}}^2)$, with defined noise scales for sampling:

\begin{equation}
\left(\sigma_{\max}^{1/\rho} + \frac{i}{N-1}\left(\sigma_{\min}^{1/\rho} - \sigma_{\max}^{1/\rho}\right)\right)^{\rho}.
\end{equation}

For our experiments, we used the same default hyperparameters as \cite{karras2022elucidating}.

Our proposed denoising network is shown in Fig. \ref{fig:architecture}. We leverage the EDGE \cite{tseng2023edge} architecture, a transformer-based decoder that incorporates timestep and music conditioning with feature-wise linear modulation (FiLM) \cite{perez2018film}. For the audio condition, while mel-spectrogram features are commonly used in music transcription, intermediate features from MFMs may provide complementary information \cite{liao2024music}. We leverage MERT \cite{li2023mert} for the MFM, and extract both mel spectrogram and intermediate MFM features for the encoder. Importantly, we apply two forms of dropout to both spectrogram and MFM features. To enable inpainting, we apply partial dropout, where a random, contiguous subsequence of the features are dropped. Moreover, we apply complete dropout, where the entire feature is dropped, to facilitate unconditional generation. To avoid confusion between the absence of noise and dropped-out features, we replace dropped-out regions with a learned null embedding. Moreover, we observed that applying a higher dropout rate to the spectrogram features is important to prevent overfitting. The features are then processed through separate transformer layers before being injected into the decoder to generate the denoised output. This output can be further refined by the reintroduction of Gaussian noise with a lower variance ($\text{\textbf{x}}_t\sim q(\text{\textbf{x}}_{t'}|\text{\textbf{x}}_t)$), and subsequent denoising at a lower timestep.



\section{Experiments}
\label{sec:experiment}

\begin{table*}[ht]
\centering
\scalebox{0.91}{
\begin{tabular}{lcccccccc}
 \Xhline{3\arrayrulewidth}
Model          & Training data                & \multicolumn{1}{l}{Model Type} & \multicolumn{1}{l}{Architecture} & Velocity & E-GMD & E-GMD (vel) & MDB  & IDMT   \\ \hline
OaF Drums \cite{callender2020improving}     & E-GMD                        & D                              & CRNN                             & $\checkmark$        & 83.40$^*$ & 61.70$^*$       & 75.19$^\dagger$ & 85.72$^*$  \\
DT-Ensemble \cite{vogl2018towards}    & TMIDT(-Bal), MDB, ENST, RBMA & D                              & CRNN                             & $\times$        & 64.98$^*$ & -           & - & 91.49$^*$      \\
ADTOF \cite{zehren2021adtof}         & ADTOF                        & D                              & CRNN                             & $\times$        & 44.95$^\dagger$ & -           & \underline{86.87$^\dagger$}  & \underline{94.67$^\dagger$} \\
hFT-Transformer \cite{toyama2023automatic}           & E-GMD                        & D                              & T                      & $\checkmark$        & 86.31 & 80.16       & 70.81 & 70.61  \\ \hline
N2N   (1-step) & E-GMD                        & G                              & T                      & $\checkmark$        & 88.00 & 79.10       & 82.26 & 91.22  \\
N2N   (5-step) & E-GMD                        & G                              & T                     & $\checkmark$        & \underline{89.24} & \underline{82.56}       & 86.66 & 93.74 \\
N2N   (10-step) & E-GMD                        & G                              & T                      & $\checkmark$        & \textbf{89.68} & \textbf{82.80}       & \textbf{87.86} & \textbf{94.90}  \\  \Xhline{3\arrayrulewidth}
\end{tabular}}
\caption{\textbf{F1 evaluation scores on various ADT benchmarks.} ADTOF and DT-Ensemble do not predict velocity. DT-Ensemble uses MDB as part of training dataset and is therefore not evaluated on MDB. $^*$Results taken from original paper. $^\dagger$Results obtained by evaluating the publicly available trained model. The best performance is shown in bold and the second best is underlined. D=Discriminative. G=Generative. T=Transformer. CRNN=Convolutional Recurrent Neural Network.}
\label{table:evaluation}
\end{table*}

\subsection{Implementation details}

\textbf{Experimental settings.} We trained our models using 4 NVIDIA A100 GPUs for $100$ epochs, with a learning rate of $3 \times 10^{-4}$
 and batch size of $64$, taking $\approx1$ day to complete. Besides using MERT as the audio encoder, we largely followed the same model settings as \cite{tseng2023edge}. For log mel-spectrogram conditioning, to enable fair comparisons, we followed previous methods \cite{callender2020improving,toyama2023automatic}, using an audio sample rate of 44.1 kHz and hop length of $441$, resulting in $10$ ms frames and 128 bins. For the MFM, we extracted intermediate (layer $10$) features from MERT 330M model. We applied complete dropout with $p=0.15$ and $p=0.30$ for MERT and spectrogram features, respectively, and partial dropout with $p=0.5$ for both. 

\vspace{5pt}

\noindent \textbf{Datasets.} For training, we used the train partition of the Expanded Groove MIDI Dataset (E-GMD) \cite{callender2020improving}, containing 440 hours of human drum performances from 43 drum kits and the only publicly available dataset with associated onset and velocity MIDI annotations. For evaluation, we used the E-GMD test set, as well as the IDMT-SMT-Drums (IDMT) \cite{dittmar2014real} and Medley DB Drums (MDB) \cite{southall2017mdb} datasets. We used the same drum hit settings as \cite{callender2020improving}, predicting 7 drum components (kick (KD), snare (SN), tom (TT), hi-hat (HH), crash cymbal (CY), ride cymbal (RD) and bell (BE)). For MDB, we remapped drum outputs to a 5-hit setting \cite{zehren2021adtof}, and for IDMT we used a 3-hit setting \cite{callender2020improving}.

\vspace{5pt}

\noindent \textbf{Evaluation metrics.} For evaluation, we used the music information retrieval library $\texttt{mir\_eval}$ \cite{raffel2014mir_eval} to compute note-wise onset and velocity transcription F1 scores. We followed established settings \cite{wu2018review,zehren2021adtof,callender2020improving}, assigning predicted note durations as $100$ms and used a tolerance of $50$ms. 

\subsection{Evaluation}
Benchmarking ADT methods is challenging due to differences in training datasets, pre- and post-processing methods, and transcription output (both in terms of the number of drum components and whether or not velocity is also estimated). The main method we compared is OaF Drums \cite{callender2020improving} because it is the only ADT method that predicts both onset and velocity. We therefore followed OaF Drums settings and also trained hFT-Transformer \cite{toyama2023automatic}, a transformer-based method developed for joint piano onset and velocity prediction, as an additional baseline to compare with more recent transformer methods. To adapt hFT-Transformer for drum transcription, we removed offset and frame branches. The performances on the E-GMD, IDMT and MDB datasets are shown in Table \ref{table:evaluation}. 

Firstly, we observed that none of the previous methods achieves strong performance across datasets. While OaF Drums significantly outperforms ADTOF on the E-GMD dataset ($83.40 > 44.95$), its performance is significantly worse on the IDMT ($85.72 < 94.67$) and MDB ($75.19 < 86.87$) datasets. Moreover, hFT-Transformer outperforms OaF Drums on the E-GMD dataset for both onset ($86.31 > 83.40$) and velocity ($80.16 > 61.70$), but its performance is worse on external data (IDMT: $70.61 < 85.72$, MDB: $70.81 < 75.19$). This suggests that previous methods are unable to generalize to drum audio from other datasets, and we identified the reliance on mel-spectrogram features as a possible reason (Sec. \ref{sec:ablation}). In contrast, our proposed method (N2N), which leverages both mel-spectrogram and MFM features, achieves robust performance across datasets. Diffusion models usually require a large number of sampling steps to achieve reasonable performance, but even at 5 sampling steps, N2N achieves state-of-the-art results on the E-GMD dataset. The strong, few-step performance may be due to the unique nature of transcription tasks, where the audio features provide a dense conditioning signal that is sufficient for accurate transcription without the need for significant refinement. Nonetheless, we observed further improvements by increasing the sampling steps, and at 10 sampling steps, N2N achieves state-of-the-art performance across all evaluated benchmarks.

\begin{figure}
  \centering
   \includegraphics[width=1.0\linewidth]{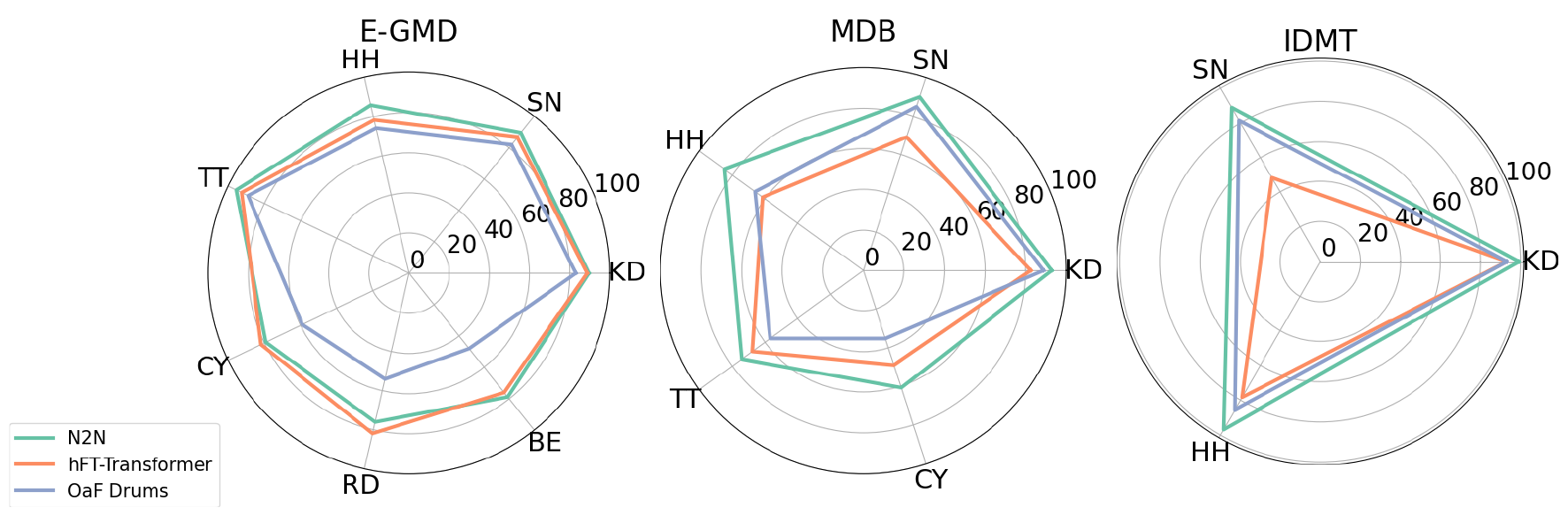}
   \caption{\textbf{Onset F1 scores per drum component}. E-GMD dataset is used for training while IDMT and MDB are external data. There are a different number of drum components labeled for each dataset, and predictions are remapped to 7, 5 and 3 components for E-GMD, MDB and IDMT datasets, respectively.}
   \label{fig:component}
\end{figure}

It is known that certain drum components, namely cymbals, are particularly challenging to predict \cite{wu2018review,callender2020improving}. We investigated the performance of different methods for predicting individual drum components (Fig. \ref{fig:component}). Across methods, we observed better performance for kick and snare drums compared to hi-hat and cymbals. This is expected given the larger range of sounds for hi-hats (which combines open and closed hi-hats) and cymbals. Compared to OaF Drums, we observed significantly better performance on these challenging classes with hFT-Transformer and N2N. However, hFT-Transformer demonstrates considerably worse performance on the IDMT and MDB datasets, most noticeably with snare drum predictions, which may reflect spectral differences between drums used across datasets. In particular, the IDMT dataset uses drum sounds from acoustic sets, sample libraries and synthesizers, which differ considerably from the E-GMD dataset used for training. In contrast, N2N achieves robust performance across datasets, with noticeable improvements for challenging drum components.

\begin{figure}
  \centering
   \includegraphics[width=0.88\linewidth]{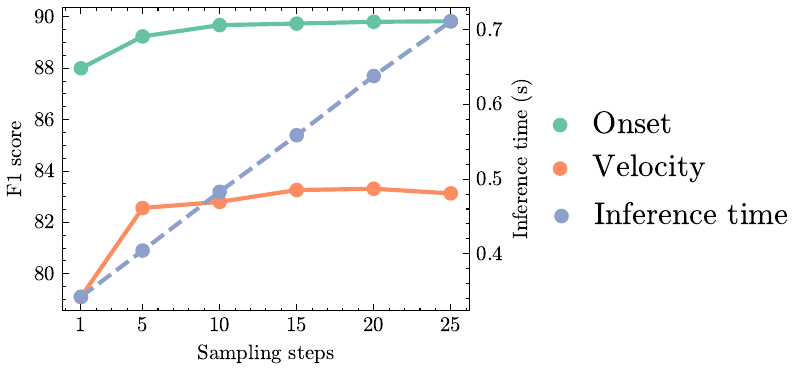}
   \caption{\textbf{Speed-accuracy trade-off}. E-GMD test set performance and associated inference time using N2N with different sampling steps. Inference time is for a single five second transcript on 1 A100 NVIDIA GPU.}
   \label{fig:speed}
\end{figure}

Unlike discriminative models, N2N uses a diffusion process that enables progressive refinement of transcriptions (Fig. \ref{fig:speed}). We observe the most dramatic performance improvements increasing one-step to few-step sampling. After around 10 sampling steps, we observe performance saturation. This suggests that the most significant performance improvements can be achieved with only a small increase in the inference costs, although it should be noted that there is an inference gap with discriminative models even at one-step (N2N: 0.163 s, hFT-Transformer: 0.086 s), due to the larger model size ($50$M$>$$5.5$M) and additional MFM feature extraction.

\begin{table}
\begin{tabular}{cccccc}
 \Xhline{3\arrayrulewidth}
Feature & Loss & E-GMD & E-GMD (vel) & MDB  & IDMT   \\ \hline
S       & MSE  & 85.55 & 66.14       & 69.20 & 80.00 \\
S       & PH   & 80.11 & 68.60       & 55.07 & 77.83   \\
S       & APH  & 86.77 & 76.17 & 71.15  & 80.89  \\
M       & APH  & \underline{87.16} & \underline{77.62}       & \underline{82.16} & \underline{90.36}  \\
S+M     & APH  & \textbf{88.00} & \textbf{79.10}       & \textbf{82.26} & \textbf{91.22}  \\  \Xhline{3\arrayrulewidth}
\end{tabular}
\caption{\textbf{Ablation study}. F1 scores for different ablation experiments using N2N. All results were obtained with one-step sampling. The best performance is shown in bold and the second best is underlined. S=Spectrogram. M=Music foundation model, MSE=Mean squared error. PH=Pseudo-Huber. APH=Annealed Pseudo-Huber.}
\label{table:ablation}
\end{table}

\begin{figure}
  \centering
    \begin{minipage}{0.49\columnwidth}
        \centering
        \includegraphics[width=\textwidth]{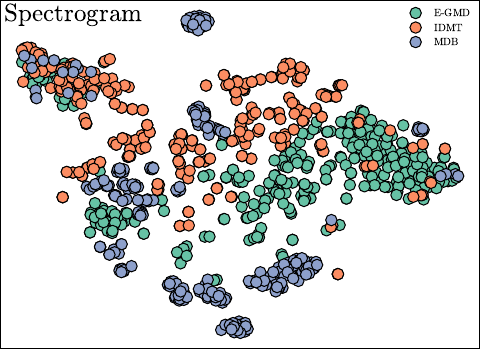}  
    \end{minipage}
    \begin{minipage}{0.49\columnwidth}
        \centering
        \includegraphics[width=\textwidth]{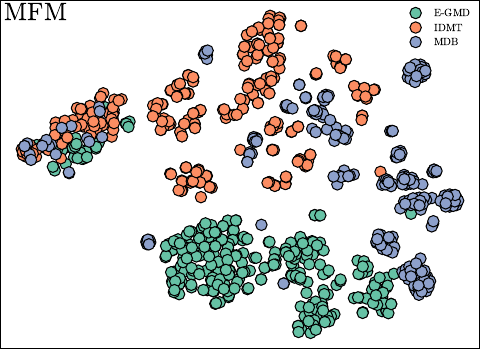} 
    \end{minipage}
   \caption{\textbf{t-SNE plots of spectrogram (left) and MFM (right) features}. Features were extracted from N2N input projection layer.}
   \label{fig:tsne}
\end{figure}

\subsection{Ablation study}
\label{sec:ablation}

We performed an ablation study for our proposed methods (Table \ref{table:ablation}). We observed that the standard MSE loss function compromises velocity prediction performance, due to onset errors dominating the overall loss, exacerbated by the squared error penalization. Using the Pseudo-Huber loss improved velocity ($68.60>66.14$) but at the cost of onset ($80.11<85.55$). Using our proposed APH loss, there is a significant improvement in velocity ($76.17>66.14$), and even a moderate improvement to the onset ($86.77>85.55$). We also investigated the effect of different audio features. While performance on in-domain E-GMD is similar, we observed dramatic improvements on the MDB ($82.16>71.15$) and IDMT ($90.36>80.89$) datasets using MFM features. To understand why conditioning on MFM features improves performance on out-of-domain data, we visualized spectrogram and MFM features from different drum datasets using a two-dimensional t-SNE \cite{maaten2008visualizing} (Fig. \ref{fig:tsne}). The MFM features between datasets are more distinct, supporting that these features contain high-level semantic information. This information can be useful to distinguish drum instruments from different domains that may have similar spectral characteristics. In fact, we observed that MFM features are complementary to spectrogram features, and obtained the best performance when both are used.

\begin{figure}[h]
  \centering
   \includegraphics[width=0.97\linewidth]{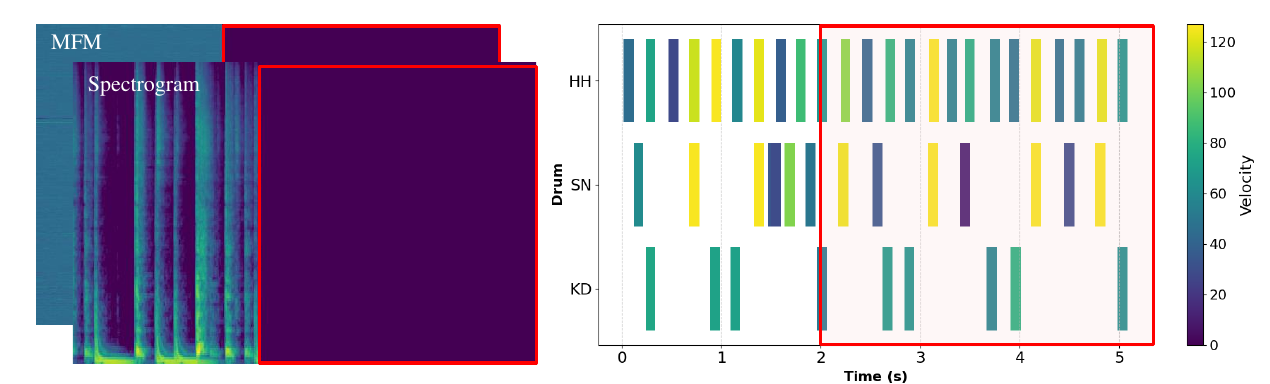}
   \caption{\textbf{Inpainting example}. Given a five-second clip where the last three seconds are masked (replaced with null embedding), N2N uses available context to generate consistent transcriptions.}
   \label{fig:inpaint}
\end{figure}

\subsection{Generative Capability}
Beyond refinement, N2N enables the generation of novel drum transcripts in the absence of audio data (Fig. \ref{fig:inpaint}). Importantly, the example demonstrates that the model leverages available audio context to create a consistent transcription, and this is the result of introducing partial dropout during training, where the model is trained to generate complete transcriptions given a partial context.

\section{Conclusion}
\label{sec:conclusion}
We reframed automatic drum transcription (ADT) as a generative task and proposed Noise-to-Notes (N2N), a diffusion model capable of transcribing drum onset and velocities from audio. We showed that using the standard diffusion objective is unable to optimize for both onset and velocity, and instead proposed an Annealed Pseudo-Huber loss that enables effective joint optimization. Moreover, we demonstrated the benefit of incorporating MFM features, in addition to commonly used spectrogram features, especially for improving transcription robustness on out-of-domain drum audio. Finally, we illustrated that reframing ADT into a generative task using diffusion models enables new inpainting and refinement capabilities. N2N sets a new state-of-the-art performance across multiple ADT benchmarks, and to our knowledge, establishes for the first time that generative models can surpass the performance of discriminative models for automatic music transcription. In future work, we will investigate methods to bridge the inference gap to discriminative models, for example leveraging distillation to reduce model size and consistency modeling to improve sampling efficiency. Additionally, we aim to investigate extending this approach to address multi-instrument transcription.




\clearpage
\newpage


\bibliographystyle{IEEE}
\bibliography{main}

@article{wu2018review,
  title={A review of automatic drum transcription},
  author={Wu, Chih-Wei and Dittmar, Christian and Southall, Carl and Vogl, Richard and Widmer, Gerhard and Hockman, Jason and M{\"u}ller, Meinard and Lerch, Alexander},
  journal={IEEE/ACM Transactions on Audio, Speech, and Language Processing},
  volume={26},
  number={9},
  pages={1457--1483},
  year={2018},
  publisher={IEEE}
}

@article{zehren2021adtof,
  title={{ADTOF}: A large dataset of non-synthetic music for automatic drum transcription},
  author={Zehren, Micka{\"e}l and Alunno, Marco and Bientinesi, Paolo},
  journal={Proc. of the 22nd Int. Society for Music Information Retrieval Conf.},
  year={2021}
}

@inproceedings{wei2021improving,
  title={Improving automatic drum transcription using large-scale audio-to-midi aligned data},
  author={Wei, I-Chieh and Wu, Chih-Wei and Su, Li},
  booktitle={ICASSP 2021-2021 IEEE International Conference on Acoustics, Speech and Signal Processing (ICASSP)},
  pages={246--250},
  year={2021},
  organization={IEEE}
}

@article{callender2020improving,
  title={Improving perceptual quality of drum transcription with the expanded groove MIDI dataset},
  author={Callender, Lee and Hawthorne, Curtis and Engel, Jesse},
  journal={arXiv preprint arXiv:2004.00188},
  year={2020}
}

@inproceedings{vogl2018towards,
    author = {Vogl, Richard and Widmer, Gerhard and Knees, Peter},
    title = {Towards multi-instrument drum transcription},
    booktitle={DAFx},
    year={2018}
}

@article{zehren2023high,
  title={High-quality and reproducible automatic drum transcription from crowdsourced data},
  author={Zehren, Micka{\"e}l and Alunno, Marco and Bientinesi, Paolo},
  journal={Signals},
  volume={4},
  number={4},
  pages={768--787},
  year={2023},
  publisher={MDPI}
}

@inproceedings{southall2017mdb,
  title={{MDB Drums}: An annotated subset of MedleyDB for automatic drum transcription},
  author={Southall, Carl and Wu, Chih-Wei and Lerch, Alexander and Hockman, Jason},
  booktitle={Late Breaking/Demos of the 18th International Society for Music Information Retrieval Conference (ISMIR)},
  year={2017},
}

@INPROCEEDINGS{10889579,
  author={Kim, Hounsu and Kwon, Taegyun and Nam, Juhan},
  booktitle={ICASSP 2025 - 2025 IEEE International Conference on Acoustics, Speech and Signal Processing (ICASSP)}, 
  title={{D3RM}: A Discrete Denoising Diffusion Refinement Model for Piano Transcription}, 
  year={2025},
  volume={},
  number={},
  pages={1-5},
  doi={10.1109/ICASSP49660.2025.10889579}}

@inproceedings{cheuk2023diffroll,
  title={{DiffRoll}: Diffusion-based generative music transcription with unsupervised pretraining capability},
  author={Cheuk, Kin Wai and Sawata, Ryosuke and Uesaka, Toshimitsu and Murata, Naoki and Takahashi, Naoya and Takahashi, Shusuke and Herremans, Dorien and Mitsufuji, Yuki},
  booktitle={ICASSP 2023-2023 IEEE International Conference on Acoustics, Speech and Signal Processing (ICASSP)},
  pages={1--5},
  year={2023},
  organization={IEEE}
}

@inproceedings{toyama2023automatic,
    author={Keisuke Toyama and Taketo Akama and Yukara Ikemiya and Yuhta Takida and Wei-Hsiang Liao and Yuki Mitsufuji},
    title={Automatic Piano Transcription with Hierarchical Frequency-Time Transformer},
    booktitle={Proceedings of the 24th International Society for Music Information Retrieval Conference},
    year={2023}
}

@inproceedings{yan2024scoring,
  title={Scoring time intervals using non-hierarchical transformer for automatic piano transcription},
  author={Yan, Yujia and Duan, Zhiyao},
  booktitle={ISMIR},
  pages={973--980},
  year={2024}
}

@ARTICLE{10896775,
  author={Wang, Qi and Liu, Mingkuan and Jia, Maoshen},
  journal={IEEE Transactions on Audio, Speech and Language Processing}, 
  title={Multitask Frequency-Time Feature-Wise Linear Modulation for Piano Transcription With Pedal}, 
  year={2025},
  volume={33},
  number={},
  pages={1049-1062},
  doi={10.1109/TASLPRO.2025.3544090}}

@inproceedings{vogl2017drum,
  title={Drum Transcription via Joint Beat and Drum Modeling Using Convolutional Recurrent Neural Networks},
  author={Vogl, Richard and Dorfer, Matthias and Widmer, Gerhard and Knees, Peter},
  booktitle={ISMIR},
  pages={150--157},
  year={2017}
}

@article{ho2020denoising,
  title={Denoising diffusion probabilistic models},
  author={Ho, Jonathan and Jain, Ajay and Abbeel, Pieter},
  journal={Advances in neural information processing systems},
  volume={33},
  pages={6840--6851},
  year={2020}
}

@article{song2019generative,
  title={Generative modeling by estimating gradients of the data distribution},
  author={Song, Yang and Ermon, Stefano},
  journal={Advances in neural information processing systems},
  volume={32},
  year={2019}
}

@inproceedings{tseng2023edge,
  title={{EDGE}: Editable dance generation from music},
  author={Tseng, Jonathan and Castellon, Rodrigo and Liu, Karen},
  booktitle={Proceedings of the IEEE/CVF conference on computer vision and pattern recognition},
  pages={448--458},
  year={2023}
}

@article{karras2022elucidating,
  title={Elucidating the design space of diffusion-based generative models},
  author={Karras, Tero and Aittala, Miika and Aila, Timo and Laine, Samuli},
  journal={Advances in neural information processing systems},
  volume={35},
  pages={26565--26577},
  year={2022}
}

@article{
    liao2024music,
    title={Music Foundation Model as Generic Booster for Music Downstream Tasks},
    author={Wei-Hsiang Liao and Yuhta Takida and Yukara Ikemiya and Zhi Zhong and Chieh-Hsin Lai and Giorgio Fabbro and Kazuki Shimada and Keisuke Toyama and Kin Wai Cheuk and Marco A. Mart{\'\i}nez-Ram{\'\i}rez and Shusuke Takahashi and Stefan Uhlich and Taketo Akama and Woosung Choi and Yuichiro Koyama and Yuki Mitsufuji},
    journal={Transactions on Machine Learning Research},
    issn={2835-8856},
    year={2025},
    url={https://openreview.net/forum?id=kHl4JzyNzF},
    note={}
    }

@inproceedings{li2023mert,
  author={Yizhi Li and Ruibin Yuan and Ge Zhang and Yinghao Ma and Xingran Chen and Hanzhi Yin and Chenghao Xiao and Chenghua Lin and Anton Ragni and Emmanouil Benetos and Norbert Gyenge and Roger B. Dannenberg and Ruibo Liu and Wenhu Chen and Gus Xia and Yemin Shi and Wenhao Huang and Zili Wang and Yike Guo and Jie Fu},
  title={{MERT}: Acoustic Music Understanding Model with Large-Scale Self-supervised Training},
  year={2024},
  booktitle={ICLR},
}

@article{charbonnier1997deterministic,
  title={Deterministic edge-preserving regularization in computed imaging},
  author={Charbonnier, Pierre and Blanc-F{\'e}raud, Laure and Aubert, Gilles and Barlaud, Michel},
  journal={IEEE Transactions on image processing},
  volume={6},
  number={2},
  pages={298--311},
  year={1997},
  publisher={IEEE}
}

@inproceedings{dittmar2014real,
  title={Real-Time Transcription and Separation of Drum Recordings Based on {NMF} Decomposition.},
  author={Dittmar, Christian and G{\"a}rtner, Daniel},
  booktitle={DAFx},
  pages={187--194},
  year={2014}
}

@inproceedings{song2024improved,
title={Improved Techniques for Training Consistency Models},
author={Yang Song and Prafulla Dhariwal},
booktitle={The Twelfth International Conference on Learning Representations},
year={2024},
url={https://openreview.net/forum?id=WNzy9bRDvG}
}

@inproceedings{perez2018film,
  title={{FiLM}: Visual reasoning with a general conditioning layer},
  author={Perez, Ethan and Strub, Florian and De Vries, Harm and Dumoulin, Vincent and Courville, Aaron},
  booktitle={Proceedings of the AAAI conference on artificial intelligence},
  volume={32},
  year={2018}
}

@inproceedings{raffel2014mir_eval,
  title={{MIR\_EVAL}: A Transparent Implementation of Common {MIR} Metrics.},
  author={Raffel, Colin and McFee, Brian and Humphrey, Eric J and Salamon, Justin and Nieto, Oriol and Liang, Dawen and Ellis, Daniel PW and Raffel, C Colin},
  booktitle={ISMIR},
  volume={10},
  pages={2014},
  year={2014}
}

@article{weber2025star,
  title={{STAR Drums}: A Dataset for Automatic Drum Transcription},
  author={Weber, Philipp and Uhle, Christian and M{\"u}ller, Meinard and Lang, Matthias},
  journal={Transactions of the International Society for Music Information Retrieval},
  volume={8},
  number={1},
  year={2025}
}

@article{maaten2008visualizing,
  title={Visualizing data using t-{SNE}},
  author={Maaten, Laurens van der and Hinton, Geoffrey},
  journal={Journal of machine learning research},
  volume={9},
  number={Nov},
  pages={2579--2605},
  year={2008}
}

@article{jaini2023intriguing,
  title={Intriguing properties of generative classifiers},
  author={Jaini, Priyank and Clark, Kevin and Geirhos, Robert},
  journal={Proceedings of the 12th International Conference on Learning Representations (ICLR)},
  year={2024}
}

\end{document}